\documentclass[journal]{IEEEtran}
\usepackage{amsmath,amssymb,amsfonts}
\usepackage{algorithmic}
\usepackage{graphicx}
\usepackage{subfigure}
\usepackage{booktabs}
\usepackage{float}
\usepackage{multirow}
\usepackage{ntheorem}
\usepackage{cite}
\usepackage{multirow}
\usepackage{setspace}
\usepackage{array}
\usepackage{color}
\usepackage[ruled]{algorithm2e}
\usepackage{hyperref}
\hypersetup{hidelinks}

\begin{document}
\bibliographystyle{IEEEtran}
\title{Blind Diagnosis for Millimeter-wave \\ Large-scale Antenna Systems}
\author{
	Rui Sun,
	Weidong Wang,
	Li Chen,
	Guo Wei, and
	Wenyi Zhang, \IEEEmembership{Senior Member, IEEE}
	
	\thanks{The authors are with the CAS Key Laboratory of Wireless-Optical Communications, University of Science and Technology of China, Hefei 230027, China (e-mail: ruisun@mail.ustc.edu.cn; wdwang@ustc.edu.cn; chenli87@ustc.edu.cn; wei@ustc.edu.cn; wenyizha@ustc.edu.cn).}
}
\maketitle

\begin{abstract}
Millimeter-wave (mmWave) communication systems rely on large-scale antenna arrays to combat large path-loss at mmWave band.
Due to hardware characteristics and deployment environments, mmWave large-scale antenna systems are vulnerable to antenna element blockages and failures, which necessitate diagnostic techniques to locate faulty antenna elements for calibration purposes.
Current diagnostic techniques require full or partial knowledge of channel state information (CSI), which can be challenging to acquire in the presence of antenna failures.
In this letter, we propose a blind diagnostic technique to identify faulty antenna elements in mmWave large-scale antenna systems, which does not require any CSI knowledge.
By jointly exploiting the sparsity of mmWave channel and failure pattern, we first formulate the diagnosis problem as a joint sparse recovery problem.
Then, the atomic norm is introduced to induce the sparsity of mmWave channel over continuous Fourier dictionary.
An efficient algorithm based on alternating direction method of multipliers (ADMM) is proposed to solve the formulated problem.
Finally, the performance of the proposed technique is evaluated through numerical simulations.
\end{abstract}


\section{Introduction}
Millimeter-wave (mmWave) large-scale antenna system is a key technology in current and next-generation mobile communication systems.
It may suffer from antenna element failures due to hardware characteristics and deploying environments.
MmWave active devices like amplifiers and mixers are generally less reliable than conventional sub-6G devices due to higher operating frequency and lower power efficiency \cite{fischer2008temperature, chang2010reliability}.
Besides, mmWave antenna elements are susceptible to blockages with comparable sizes like dirt and precipitation owing to the short wavelength at the mmWave band \cite{vinyaikin1994attenuation}.
The existence of antenna element failure will cause signal power reduction and radiation pattern distortion, which may lead to severe degradation in system performance.
Therefore, diagnostic techniques for mmWave large-scale antenna systems are of significant interest for system monitoring and maintenance.

Extensive related works on antenna array diagnosis have been proposed, including compressed sensing based techniques \cite{migliore2011compressed, eltayeb2018compressive, palmeri2019diagnosis, xiong2019compressed, ince2015array, salucci2018planar, ma2020antenna} and deep learning based techniques \cite{chen2019deep}.
The main idea of compressed sensing based diagnostic techniques is to use known channel state information (CSI) to generate fault-free reference signal and then subtract it from the received signal.
Therefore, the differential signal contains the information of faulty antenna elements, which can then be estimated from compressed measurements.
The work \cite{migliore2011compressed} is the first to introduce compressed sensing to array diagnosis.
Improvements over \cite{migliore2011compressed} include modifications in sampling methods \cite{eltayeb2018compressive, palmeri2019diagnosis, xiong2019compressed} and refinements in sparse recovery algorithms \cite{ince2015array, salucci2018planar, ma2020antenna}.
In particular, the work in \cite{sun2021hybrid} extended antenna-only diagnosis to joint diagnosis of antenna, phase shifter, and RF chain for hybrid beamforming (HBF) systems.
Deep-learning based diagnostic techniques include \cite{chen2019deep}, which utilizes a convolutional neural network (CNN) to detect the abnormality in the distribution of received signal and then locate faulty antenna elements.

In order to distinguish antenna failure from channel fading, the above diagnostic techniques rely on perfect CSI to generate fault-free reference signal, which can be challenging to acquire in the presence of antenna failures.
Recently, a diagnostic technique proposed in \cite{medina2020millimeter} relaxed the CSI requirement, only requiring the angle-of-arrival (AOA) of each sub-path in the channel.
However, acquiring the AOA of each sub-path can still be a challenging task for a potentially faulty system since faulty antenna elements will change the array geometry and distort the radiation pattern, rendering AOA estimation highly unreliable.

Aiming at this limitation, in this letter, we propose a blind diagnostic technique for mmWave large-scale antenna systems.
The term `blind' indicates that the proposed technique does not require any knowledge of the CSI.
This is achieved by jointly exploiting the failure pattern sparsity and the mmWave channel sparsity \cite{song2018scalable, song2019efficient, zhang2019codebook}.
Specifically, we first formulate the diagnosis problem as a joint sparse recovery problem.
Then, the atomic norm is introduced to induce the sparsity of mmWave channel over the continuous Fourier dictionary.
An efficient algorithm based on alternating direction method of multipliers (ADMM) is proposed to solve the joint sparse recovery problem.
Finally, numerical simulations validate the proposed technique.

\textit{Notations:}
We use a lowercase and an uppercase bold letter to represent a vector and a matrix, respectively.
$\Vert \mathbf{a} \Vert_1$ and $\Vert \mathbf{a} \Vert_2$ represent the $l_1$ and $l_2$ norm of vector $\mathbf{a}$, respectively.
In particular, $\Vert \mathbf{a} \Vert_\mathcal{A}$ denotes the atomic norm of $\mathbf{a}$.
For a matrix $\mathbf{A}$, $\mathbf{A}^\mathrm{T}$, $\mathbf{A}^\mathrm{H}$ denote its transpose and conjugate transpose, respectively.
$\Vert \mathbf{A} \Vert_\mathrm{F}$ and $\mathrm{tr}(\mathbf{A})$ represent the Frobenius norm and the trace of $\mathbf{A}$, respectively.
$\mathbf{I}_N$ represents an $N \times N$ identity matrix.
$\mathcal{CN}(\mu, \sigma^2)$ denotes circularly symmetric complex Gaussian distribution with mean $\mu$ and variance $\sigma^2$.
$U(a,b)$ denotes uniform distribution over the interval $[a,b]$.

\section{System Model}
\label{sec:SysModel}
We consider an analog beamforming (ABF) mmWave large-scale antenna system equipped with an $N$-element uniform linear array (ULA)\footnote{
	The proposed diagnostic technique can be extended to other system architectures (like digital beamforming (DBF) and hybrid beamforming (HBF) systems) and other array structures (like uniform planer array (UPA)).
	For HBF systems, each RF chain can obtain independent measurements in a single time slot using different phase shifts, which shortens the measurement time by a factor of the number of RF chains.
	For DBF systems, the measurement can be accomplished in a single time slot since each antenna has its own RF chain.
}, which is the array-under-test (AUT).
A single-antenna diagnostic transmitter (TX) is adopted to transmit test symbols.
The AUT receives the test symbol and conducts the diagnosis.
In the absence of antenna failure, the fault-free received symbol can be expressed as
\begin{equation}
y' = \mathbf{f}^\mathrm{T} \mathbf{h} x + w,
\end{equation}
where $\mathbf{f} \in \mathbb{C}^N$ is the combining vector, $\mathbf{h} \in \mathbb{C}^N$ is the channel vector between the TX and the AUT, $x$ is the transmitted symbol, $w \sim \mathcal{CN}(0,1/\mathrm{SNR})$ is the normalized noise and $\mathrm{SNR}$ is the signal-to-noise ratio.

The channel $\mathbf{h}$ is assumed as a block-fading mmWave clustered channel \cite{zhang2019codebook}, which can be expressed as
\begin{equation}
\label{chanModel}
\mathbf{h} = \sum_{l=1}^{L} \alpha_l \mathbf{a}(\theta_l),
\end{equation}
where
$
\mathbf{a}(\theta_l) = [1, e^{j 2\pi d \sin \theta_l}, \cdots, e^{j 2\pi d (N-1) \sin \theta_l}]^\mathrm{T}
$
is the array response vector, $L$ is the number of sub-paths, $\alpha_l \sim \mathcal{CN}(0, 1/L)$ and $\theta_l \sim U(-\pi/2, \pi/2)$ are the complex gain and the angle-of-arrival (AOA) of the $l$-th sub-path, respectively, and $d=1/2$ is the element spacing relative to the wavelength.

In the presence of faulty antenna elements, the actual channel vector will deviate from the ideal one since faults may cause additional attenuation and phase shift to the channel response.
Therefore, the received symbol under antenna faults can be expressed as
\begin{equation}
y = \mathbf{f}^\mathrm{T} (\mathbf{h} + \mathbf{h}_\mathrm{f}) x + w,
\end{equation}
where $\mathbf{h}_\mathrm{f} \in \mathbb{C}^N$ is the fault-induced channel deviation.
Due to the \textit{failure pattern sparsity} that usually only a small number of antenna elements are faulty \cite{migliore2011compressed, eltayeb2018compressive}, $\mathbf{h}_\mathrm{f}$ is assumed as a sparse vector, in which non-zero entries indicate faulty antenna elements.

To detect faults, the TX transmits test symbol $x=1$ throughout the diagnosis.
Within the channel coherence time, the AUT uses $K$ random combining vectors to receive the signal, yielding the observation model
\begin{equation}
\label{obserModel}
\mathbf{y} = \mathbf{F}(\mathbf{h} + \mathbf{h}_\mathrm{f}) + \mathbf{w},
\end{equation}
where
\begin{equation}
\begin{aligned}
\mathbf{y} &= [y_1, \cdots, y_K]^\mathrm{T} \in \mathbb{C}^K, \\
\mathbf{F} &= [\mathbf{f}_1, \cdots, \mathbf{f}_K]^\mathrm{T} \in \mathbb{C}^{K \times N}, \\
\mathbf{w} &= [w_1, \cdots, w_K]^\mathrm{T} \in \mathbb{C}^K, \\
\end{aligned}
\end{equation}
in which $y_k$, $\mathbf{f}_k$, and $w_k$ are the received symbol, the combining vector, and the noise corresponded to the $k$-th measurement, respectively.

The goal of diagnosis is to recover the sparse fault-induced channel deviation $\mathbf{h}_\mathrm{f}$, in which non-zero entries indicate faulty antenna elements.
From \eqref{obserModel}, it can be observed that $\mathbf{h}_\mathrm{f}$ is coupled with the fault-free channel vector $\mathbf{h}$.
Since the system contains potentially faulty antenna elements, the channel vector $\mathbf{h}$ can not be estimated by conventional channel estimation techniques like pilot-based training, and thus recovering $\mathbf{h}_\mathrm{f}$ under unknown $\mathbf{h}$ can be challenging.

To cope with this issue, prior works require full or partial knowledge of the CSI to perform the diagnosis.
The diagnostic technique proposed in \cite{eltayeb2018compressive} requires full CSI, which assumes a free-space wireless channel and proposes to calculate the channel vector using known sub-path AOA and gain (i.e., $\theta_l$ and $\alpha_l$).
Hence, one can generate the fault-free channel vector $\mathbf{h}$ using \eqref{chanModel} and subtract it from the received signal, and the impact of channel can be eliminated.
A recent work \cite{medina2020millimeter} relaxes the CSI requirement in the sense that it only requires the AOA of each sub-path (i.e., $\theta_l$).
There the key idea is to project the received signal onto the null space of AOAs and thus the impact of channel can also be eliminated.

The above diagnostic techniques follow the routine of ``cancel-then-recover" in the sense that they seek to cancel the impact of channel from the received signal first and then recover the fault-induced channel deviation.
The requirement of CSI can be challenging to satisfy, especially for the outdoor online diagnosis in a complex multipath scattering environment.
In the following, we propose a diagnostic technique that does not require any knowledge of the CSI (i.e., blind diagnosis).
This is achieved by jointly exploiting the mmWave channel sparsity and the failure pattern sparsity.
Besides, the fault-free channel vector $\mathbf{h}$ can also be recovered, which provides an approach to perform channel estimation under antenna element failures.

\section{Fault Diagnosis}
\subsection{Problem Formulation}
Recall that our goal is to recover $\mathbf{h}_\mathrm{f}$ under the observation model \eqref{obserModel}, in which non-zero entries indicate faulty antenna elements.
At first glance, it seems impossible to distinguish channel fading from antenna failures, and hence recovering $\mathbf{h}_\mathrm{f}$ under unknown $\mathbf{h}$ appears to be a challenging task.
To decouple the failure deviation $\mathbf{h}_\mathrm{f}$ from the channel $\mathbf{h}$, we need to exploit their structural characteristics.
The fault-induced channel deviation $\mathbf{h}_\mathrm{f}$ is sparse itself due to the failure pattern sparsity.
The ideal channel vector $\mathbf{h}$ is also sparse in the Fourier dictionary since the mmWave channel is sparse in the angle domain (i.e., $L \ll N$) \cite{song2018scalable, song2019efficient}.
Therefore, $\mathbf{h}$ and $\mathbf{h}_\mathrm{f}$ are sparse under different dictionaries, and this property allows them to be recovered simultaneously \cite{duarte2006sparse}.

To jointly recover $\mathbf{h}$ and $\mathbf{h}_\mathrm{f}$, we need to introduce \textit{a-prior} information on them.
We adopt the well-known $l_1$ norm as the sparsity-inducing norm for $\mathbf{h}_\mathrm{f}$:
\begin{equation}
\Vert \mathbf{h}_\mathrm{f} \Vert_1 = \sum_{n=1}^{N} \vert h_{\mathrm{f}, n} \vert,
\end{equation}
where $h_{\mathrm{f}, n}$ is the $n$-th entry in $\mathbf{h}_\mathrm{f}$.

The ideal channel vector $\mathbf{h}$ is sparse in the Fourier dictionary.
To be more specific, $\mathbf{h}$ is sparse in the continuous Fourier dictionary
\begin{equation}
\mathcal{A} = \left\{ \mathbf{a}(\theta_l) \vert \theta_l \in [-\pi/2, \pi/2] \right\}.
\end{equation}
We adopt the atomic norm as the sparsity-inducing norm for $\mathbf{h}$.
The atomic norm of $\mathbf{h}$ over the dictionary $\mathcal{A}$ is defined as  \cite{tang2013compressed}
\begin{equation}
\Vert \mathbf{h} \Vert_\mathcal{A} = \inf \left\{ \sum_l |\alpha_l| \bigg| \mathbf{h} = \sum_l \alpha_l \mathbf{a}_l, \mathbf{a}_l \in \mathcal{A} \right\},
\end{equation}
which can be regarded as the $l_1$ norm over the continuous dictionary $\mathcal{A}$.

Using the $l_1$ norm and the atomic norm as sparsity-inducing norm for $\mathbf{h}_\mathrm{f}$ and $\mathbf{h}$, respectively, the optimization problem for joint recovery can be expressed as
\begin{equation}
\label{opt1}
\begin{aligned}
\{ \hat{\mathbf{h}}, \hat{\mathbf{h}}_\mathrm{f} \} = \arg \min_{\mathbf{h}, \mathbf{h}_\mathrm{f}} \frac{1}{2} \Vert \mathbf{y} - \mathbf{F}(\mathbf{h} + \mathbf{h}_\mathrm{f}) \Vert_2^2 + \tau \Vert \mathbf{h} \Vert_\mathcal{A} + \lambda \Vert \mathbf{h}_\mathrm{f} \Vert_1,
\end{aligned}
\end{equation}
where $\tau$ and $\lambda$ are regularization parameters controlling the sparsity penalty on $\mathbf{h}$ and $\mathbf{h}_\mathrm{f}$, respectively.
The optimization problem \eqref{opt1} tends to find a sparse vector $\mathbf{h}_\mathrm{f}$ and a vector $\mathbf{h}$ that is sparse under the dictionary $\mathcal{A}$ to fit the observation model.

Although the optimization problem \eqref{opt1} is clear in its form, it can not be directly solved since it involves infinite-dimensional variable optimization due to the continuity of the dictionary $\mathcal{A}$.
To cope with this issue, a conventional approach is to discretize the angle domain into a grid, which forms a discrete Fourier transform (DFT) dictionary \cite{yang2012off}.
This approach, however, induces the off-grid error since the AOAs will not lie exactly on the grid, which may significantly degrades the recovery performance \cite{chi2011sensitivity}.
In the following, we develop an efficient algorithm for solving the optimization problem \eqref{opt1}.

\subsection{An Efficient Diagnostic Algorithm}
The main difficulty of solving \eqref{opt1} arises from the atomic norm $\Vert \mathbf{h} \Vert_\mathcal{A}$ over the continuous Fourier dictionary $\mathcal{A}$.
Fortunately, the minimization of $\Vert \mathbf{h} \Vert_\mathcal{A}$ admits the following semidefinite program (SDP) thanks to the Carathéodory–Fejér–Pisarenko decomposition \cite{georgiou2007caratheodory, tang2013compressed}
\begin{equation}
\label{SDPori}
\arg \min_{\substack{\mathbf{u} \in \mathbb{C}^N \\ v \in \mathbb{R}}}
\frac{1}{2}\left(\frac{1}{N} \mathrm{tr}\big(T(\mathbf{u})\big) + v \right)
\hspace{1em} \mathrm{s.t.} \hspace{1em}
\begin{bmatrix}
T(\mathbf{u}) & \mathbf{h} \\
\mathbf{h}^\mathrm{H} & v
\end{bmatrix}
\succeq 0,
\end{equation}
where $\mathrm{tr}(\cdot)$ denotes the trace of a matrix, and $T(\mathbf{u})$ is the Hermitian Toeplitz matrix with vector $\mathbf{u}$ as its first column.

Therefore, the optimization problem \eqref{opt1} has the equivalent SDP
\begin{equation}
\label{opt2}
\begin{aligned}
\{ \hat{\mathbf{h}}, \hat{\mathbf{h}}_\mathrm{f} \}
=&\arg \min_{\mathbf{h}, \mathbf{h}_\mathrm{f}, \mathbf{u}, v} \frac{1}{2} \Vert \mathbf{y} - \mathbf{F}(\mathbf{h} + \mathbf{h}_\mathrm{f}) \Vert_2^2 \\
&+ \frac{\tau}{2}\left(\frac{1}{N} \mathrm{tr}\big(T(\mathbf{u})\big) + v \right) + \lambda \Vert \mathbf{h}_\mathrm{f} \Vert_1 \\
&\mathrm{s.t.} \hspace{1em}
\begin{bmatrix}
	T(\mathbf{u}) & \mathbf{h} \\
	\mathbf{h}^\mathrm{H} & v
\end{bmatrix}
\succeq 0.
\end{aligned}
\end{equation}

Related works on the atomic norm suggest using the CVX toolbox to solve the SDP \eqref{opt2} \cite{tang2013compressed}, which can be time-consuming for large-scale problems.
To solve \eqref{opt2} in an efficient way, we develop an algorithm based on the alternating direction method of multipliers (ADMM) algorithm \cite{boyd2011distributed}.
The ADMM algorithm integrates the augmented Lagrangian method with the dual ascent method, which usually shows excellent efficiency in solving large-scale problems.

We first rewrite \eqref{opt2} into the ADMM form:
\begin{equation}
\begin{aligned}
\{ \hat{\mathbf{h}}, \hat{\mathbf{h}}_\mathrm{f} \}
=& \arg \min_{\mathbf{h}, \mathbf{h}_\mathrm{f}, \mathbf{u}, v} \frac{1}{2}\Vert \mathbf{y} - \mathbf{F}(\mathbf{h} + \mathbf{h}_\mathrm{f}) \Vert_2^2 \\
&+ \frac{\tau}{2} \left( \frac{1}{N} \mathrm{tr}\big(T(\mathbf{u})\big) + v \right)
+ \lambda \Vert \mathbf{h}_\mathrm{f} \Vert_1, \\
&\mathrm{s.t.} \hspace{1em} \mathbf{Z} = 
\begin{bmatrix}
T(\mathbf{u}) & \mathbf{h} \\
\mathbf{h}^\mathrm{H} & v
\end{bmatrix}
\succeq 0,
\end{aligned}
\label{ADMM1}
\end{equation}
where $\mathbf{Z} \in \mathbb{C}^{(N+1) \times (N+1)}$ is an auxiliary matrix.

The augmented Lagrangian function for \eqref{ADMM1} can be expressed as
\begin{equation}
\label{ADMM2}
\begin{aligned}
&\mathcal{L}_\rho(v, \mathbf{u, h}, \mathbf{h}_\mathrm{f}, \mathbf{Z}, \boldsymbol{\Lambda}) \\
&= \frac{1}{2}\Vert \mathbf{y} - \mathbf{F}(\mathbf{h} + \mathbf{h}_\mathrm{f}) \Vert_2^2
+ \frac{\tau}{2} \left( \frac{1}{N} \mathrm{tr}\big(T(\mathbf{u})\big) + v \right)
+ \lambda \Vert \mathbf{h}_\mathrm{f} \Vert_1 \\
&+ \left\langle \boldsymbol{\Lambda}, \mathbf{Z} - \begin{bmatrix}
T(\mathbf{u}) & \mathbf{h} \\
\mathbf{h}^\mathrm{H} & v
\end{bmatrix} \right \rangle
+ \frac{\rho}{2} \left\Vert \mathbf{Z} - \begin{bmatrix}
T(\mathbf{u}) & \mathbf{h} \\
\mathbf{h}^\mathrm{H} & v
\end{bmatrix} \right\Vert_\mathrm{F}^2,
\end{aligned}
\end{equation}
where $\boldsymbol{\Lambda} \in \mathbb{C}^{(N+1) \times (N+1)}$ is the Lagrangian multiplier, $\rho$ is a penalty parameter, and $\langle \cdot, \cdot \rangle$ denotes the real inner product.

The ADMM algorithm minimizes the augmented Lagrangian function by iteratively updating the following variables:
\begin{equation}
\begin{aligned}
v^{(l+1)} &= \arg \min_{v} \mathcal{L}_\rho(v, \mathbf{u}^{(l)}, \mathbf{h}^{(l)}, \mathbf{h}_\mathrm{f}^{(l)}, \mathbf{Z}^{(l)}, \boldsymbol{\Lambda}^{(l)}), \\
\mathbf{u}^{(l+1)} &= \arg \min_\mathbf{u} \mathcal{L}_\rho(v^{(l+1)}, \mathbf{u}, \mathbf{h}^{(l)}, \mathbf{h}_\mathrm{f}^{(l)}, \mathbf{Z}^{(l)}, \boldsymbol{\Lambda}^{(l)}), \\
\mathbf{h}^{(l+1)} &= \arg \min_\mathbf{h} \mathcal{L}_\rho(v^{(l+1)}, \mathbf{u}^{(l+1)}, \mathbf{h}, \mathbf{h}_\mathrm{f}^{(l)}, \mathbf{Z}^{(l)}, \boldsymbol{\Lambda}^{(l)}), \\
\mathbf{h}_\mathrm{f}^{(l+1)} &= \arg \min_{\mathbf{h}_\mathrm{f}} \mathcal{L}_\rho(v^{(l+1)}, \mathbf{u}^{(l+1)}, \mathbf{h}^{(l+1)}, \mathbf{h}_\mathrm{f}, \mathbf{Z}^{(l)}, \boldsymbol{\Lambda}^{(l)}), \\
\mathbf{Z}^{(l+1)} &= \arg \min_\mathbf{Z} \mathcal{L}_\rho(v^{(l+1)}, \mathbf{u}^{(l+1)}, \mathbf{h}^{(l+1)}, \mathbf{h}_\mathrm{f}^{(l+1)}, \mathbf{Z}, \boldsymbol{\Lambda}^{(l)}), \\
\boldsymbol{\Lambda}^{(l+1)} &= \boldsymbol{\Lambda}^{(l)} + \rho \left( \mathbf{Z}^{(l+1)} - \begin{bmatrix}
T\left(\mathbf{u}^{(l+1)}\right) & \mathbf{h}^{(l+1)} \\
{\mathbf{h}^{(l+1)}}^\mathrm{H} & v^{(l+1)}
\end{bmatrix} \right),
\end{aligned}
\end{equation}
where the superscript $(l)$ denotes the $l$-th iteration.

To perform the above updates in an explicit way, we first introduce the partitions
\begin{equation}
\begin{aligned}
\mathbf{Z}^{(l)} =
\begin{bmatrix}
\mathbf{Z}^{(l)}_0 & \mathbf{z}^{(l)}_1 \\
{\mathbf{z}^{(l)}_1}^\mathrm{H} & Z^{(l)}_{N+1, N+1}
\end{bmatrix},
\boldsymbol{\Lambda}^{(l)} =
\begin{bmatrix}
\boldsymbol{\Lambda}^{(l)}_0 & \boldsymbol{\lambda}^{(l)}_1 \\
{\boldsymbol{\lambda}^{(l)}_1}^\mathrm{H} & \Lambda^{(l)}_{N+1, N+1}
\end{bmatrix}.
\end{aligned}
\end{equation}

The augmented Lagrangian function is convex and differentiable with respect to $v$, $\mathbf{u}$, and $\mathbf{h}$.
Setting their gradients to zero, these updates have closed-form expressions as
\begin{equation}
\label{update_u}
\begin{aligned}
v^{(l+1)} &= Z^{(l)}_{N+1, N+1} + \frac{1}{\rho} \left(\Lambda^{(l)}_{N+1, N+1} - \frac{\tau}{2} \right), \\
\mathbf{u}^{(l+1)} &= \boldsymbol{\Psi}^{-1} \left( T^*\left( \mathbf{Z}^{(l)}_0 + \frac{1}{\rho} \boldsymbol{\Lambda}^{(l)}_0 \right) \right) - \frac{\tau}{2 \rho} \mathbf{e}_1, \\
\mathbf{h}^{(l+1)} &= (\mathbf{F}^\mathrm{H} \mathbf{F} + 2\rho \mathbf{I}_{N})^{-1} \left( \mathbf{F}^\mathrm{H}(\mathbf{y} - \mathbf{F} \mathbf{h}^{(l)}_\mathrm{f}) + 2\boldsymbol{\lambda}^{(l)}_1 + 2 \rho \mathbf{z}^{(l)}_1 \right),
\end{aligned}
\end{equation}
where $\boldsymbol{\Psi}$ is a diagonal matrix and $\boldsymbol{\Psi}_{i,j} = N - j + 1, j = 1, \cdots, N$, $\mathbf{e}_1$ is a zero vector with its first entry being one, and $T^*(\cdot)$ generates a vector whose $i$-th element is the trace of the $(i-1)$-th subdiagonal of the input matrix.

The update of $\mathbf{h}_\mathrm{f}$ can be written as
\begin{equation}
\begin{aligned}
\mathbf{h}_\mathrm{f}^{(l+1)}
&= \arg \min_{\mathbf{h}_\mathrm{f}} \frac{1}{2}\Vert (\mathbf{y} - \mathbf{F} \mathbf{h}^{(l+1)}) - \mathbf{F} \mathbf{h}_\mathrm{f} \Vert_2^2 + \lambda \Vert \mathbf{h}_\mathrm{f} \Vert_1,
\end{aligned}
\end{equation}
which is the well-known LASSO problem and can be solved efficiently by the algorithm proposed in \cite{boyd2011distributed}.

Finally, the update of $\mathbf{Z}$ can be expressed as
\begin{equation}
\begin{aligned}
\mathbf{Z}^{(l+1)}
&= \arg \min_{\mathbf{Z} \succeq 0} \left\Vert \mathbf{Z} - 
\left(
\underbrace{
\begin{bmatrix}
T\left(\mathbf{u}^{(l+1)}\right) & \mathbf{h}^{(l+1)} \\
{\mathbf{h}^{(l+1)}}^\mathrm{H} & v^{(l+1)}
\end{bmatrix}
- \frac{1}{\rho}\boldsymbol{\Lambda}^{(l)}
}_\mathbf{G}
\right)
\right\Vert_\mathrm{F}^2,
\end{aligned}
\end{equation}
which amounts to projecting the Hermitian matrix $\mathbf{G}$ onto the positive semidefinite cone and can be performed by computing the eigenvalue decomposition of $\mathbf{G}$ and setting all negative eigenvalues to zero.

The overall algorithm for solving \eqref{opt2} is summarized in Algorithm \ref{algo1}, where $\epsilon$ and $l_\mathrm{max}$ are the halting threshold and the maximum number of iterations\footnote{Empirical results suggest that setting $\epsilon = 10^{-3}$ and $l_\mathrm{max} = 1000$ typically leads to a good performance.}.
The computational complexity mainly arises from the updates of $\mathbf{h}$, $\mathbf{h}_\mathrm{f}$, and $\mathbf{Z}$.
The updates of $\mathbf{h}$ and $\mathbf{h}_\mathrm{f}$ involve matrix inversion, which has the complexity of $O(N^3)$.
The update of $\mathbf{Z}$ requires eigenvalue decomposition and has the complexity of $O(N^3)$.
Therefore, the overall computational complexity of Algorithm \ref{algo1} is $O(N^3)$.

\begin{algorithm}[t]
\caption{An Efficient Algorithm for Solving \eqref{opt2}}
\label{algo1}

Initialization: $\mathbf{Z}^{(0)} = \mathbf{0}$, $\boldsymbol{\Lambda}^{(0)} = \mathbf{0}$, $\mathbf{h}_\mathrm{f}^{(0)} = \mathbf{0}$; \\

\While{$\Vert \mathbf{h}_\mathrm{f}^{(l+1)} - \mathbf{h}_\mathrm{f}^{(l)}\Vert_2 > \epsilon$ and $l\le l_\mathrm{max}$}{

\nl $v^{(l+1)} = Z^{(l)}_{N+1, N+1} + \frac{1}{\rho} \left(\Lambda^{(l)}_{N+1, N+1} - \frac{\tau}{2} \right)$; \\

\nl $\mathbf{u}^{(l+1)} = \boldsymbol{\Psi}^{-1} \left( T^*\left( \mathbf{Z}^{(l)}_0 + \frac{1}{\rho} \boldsymbol{\Lambda}^{(l)}_0 \right) \right) - \frac{\tau}{2 \rho} \mathbf{e}_1$; \\
		
\nl $\mathbf{h}^{(l+1)} = (\mathbf{F}^\mathrm{H} \mathbf{F} + 2\rho \mathbf{I}_{N})^{-1} \left( \mathbf{F}^\mathrm{H}(\mathbf{y} - \mathbf{F} \mathbf{h}^{(l)}_\mathrm{f}) + 2\boldsymbol{\lambda}^{(l)}_1 + 2 \rho \mathbf{z}^{(l)}_1 \right)$; \\
		
\nl $\mathbf{h}_\mathrm{f}^{(l+1)} = \arg \min_{\mathbf{h}_\mathrm{f}} \frac{1}{2}\Vert (\mathbf{y} - \mathbf{F} \mathbf{h}^{(l+1)}) - \mathbf{F} \mathbf{h}_\mathrm{f} \Vert_2^2 + \lambda \Vert \mathbf{h}_\mathrm{f} \Vert_1$, solved by the algorithm proposed in \cite{boyd2011distributed}; \\
		
\nl Perform eigenvalue decomposition on
$
\begin{bmatrix}
T\left(\mathbf{u}^{(l+1)}\right) & \mathbf{h}^{(l+1)} \\
{\mathbf{h}^{(l+1)}}^\mathrm{H} & v^{(l+1)}
\end{bmatrix}
- \frac{1}{\rho}\boldsymbol{\Lambda}^{(l)}
$
and set all negative eigenvalues to zero, yielding $\mathbf{Z}^{(l+1)}$; \\
		
\nl $\boldsymbol{\Lambda}^{(l+1)} = \boldsymbol{\Lambda}^{(l)} + \rho \left( \mathbf{Z}^{(l+1)} - 
\begin{bmatrix}
T\left(\mathbf{u}^{(l+1)}\right) & \mathbf{h}^{(l+1)} \\
{\mathbf{h}^{(l+1)}}^\mathrm{H} & v^{(l+1)}
\end{bmatrix}
\right)$;
}
\end{algorithm}

\section{Numerical Simulations}
\label{sec:simu}
In this section, we conduct numerical simulations to evaluate the performance of the proposed diagnostic technique.
The number of antenna elements is set to $N=64$.
The number of faulty antenna elements is set to 3, whose locations are chosen uniformly at random.
The amplitude and phase of the entries in the fault-induced channel deviation $\mathbf{h}_\mathrm{f}$ follow $U(0.2, 1)$ and $U(0, 2\pi)$, respectively.
Regularization parameters are set to $\lambda = 0.4$ and $\tau=0.3$.
We adopt the success probability as the performance metric, which is defined as the probability that the states of all antenna elements are correctly identified.
For comparison, we adopt the diagnostic techniques proposed in \cite{eltayeb2018compressive} (denoted as Eltayeb18) and \cite{medina2020millimeter} (denoted as Medina20) as benchmark techniques, which require full and partial CSI, respectively.

\begin{figure}[t]
	\centering
	\includegraphics[scale=0.4]{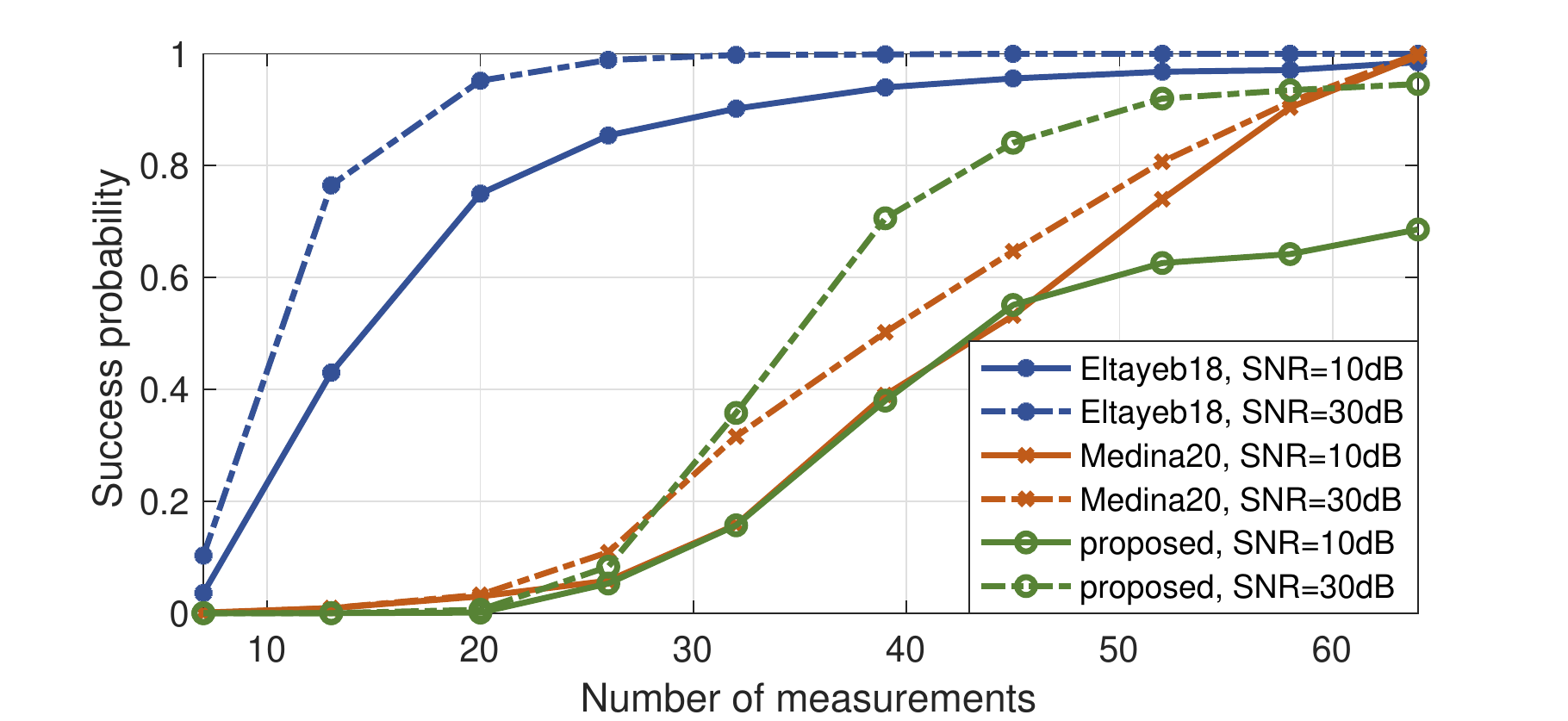}
	\caption{Success probability versus the number of measurements under different SNRs.}
	\label{F1}
\end{figure}

First, we evaluate the effect of the number of measurements.
Fig. \ref{F1} shows the success probability versus the number of measurements under different SNRs, where the number of sub-paths is set to $L=4$.
It can be observed that all techniques perform better with a larger number of measurements and/or larger SNR.
Among all diagnostic techniques, the diagnosis with full CSI (Eltayeb18) outperforms others since the impact of channel can be completely eliminated using known CSI.
When the SNR is sufficiently large, the proposed technique has slightly better performance than the diagnosis with partial CSI (Medina20).

\begin{figure}[t]
	\centering
	\includegraphics[scale=0.4]{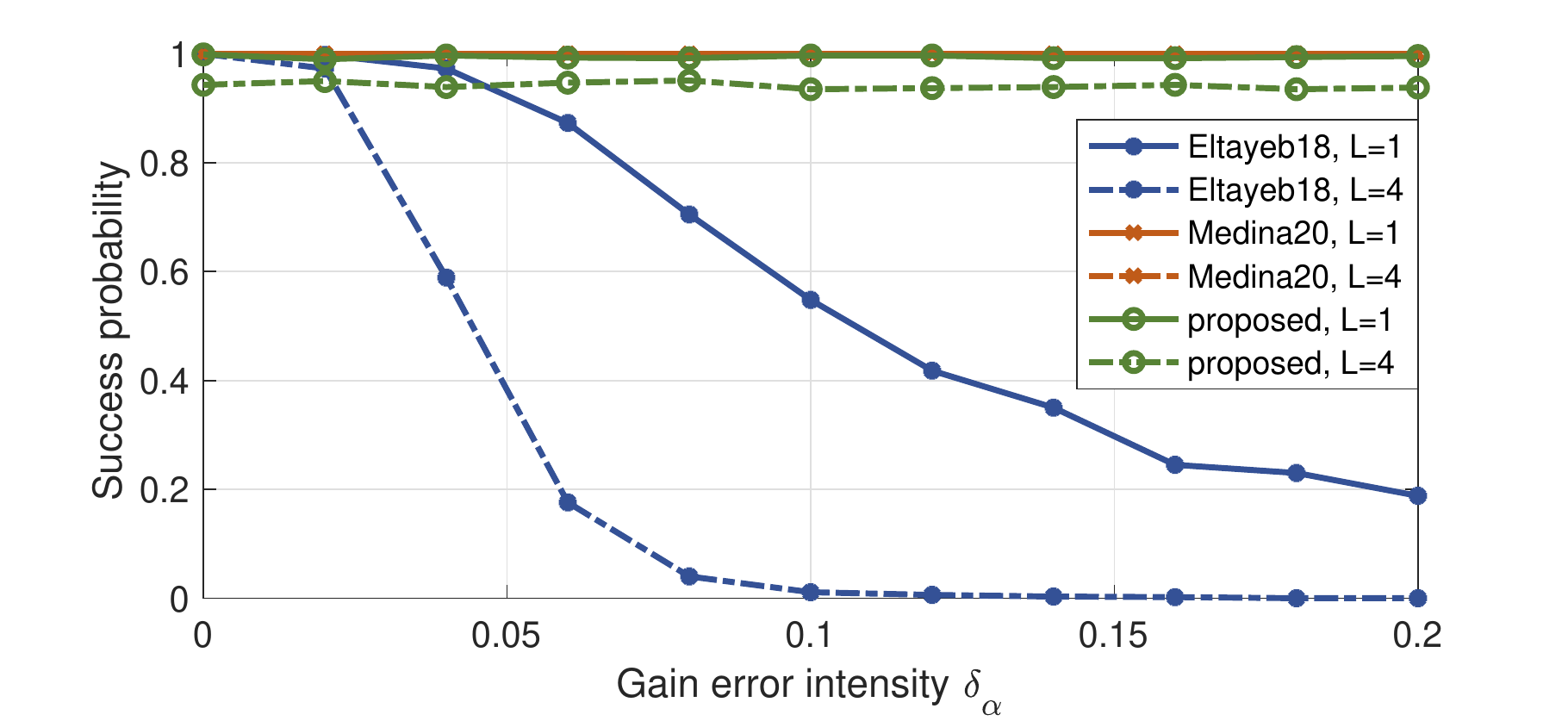}
	\caption{Success probability versus sub-path gain error intensity under different number of sub-paths.}
	\label{F2}
\end{figure}

Next, we assume that the estimated sub-path gain $\hat{\alpha}_l$ contains estimation error, which is defined as
\begin{equation}
	\hat{\alpha}_l = \alpha_l + \delta_\alpha \alpha_\mathrm{e},
\end{equation}
where $\alpha_\mathrm{e} \sim \mathcal{CN}(0, 1)$ represents gain estimation error and $\delta_\alpha$ is the error intensity.
The performance of different techniques under sub-path gain estimation error is shown in Fig. \ref{F2}, where $\mathrm{SNR}=30$dB and the number of measurements is set to $K=N$ to ensure sufficient measurements for all techniques.
We can observe that the performance of diagnosis with full CSI (Eltayeb18) degrades significantly, while other techniques are not affected by the estimation error since they do not require the knowledge of sub-path gain.
Besides, the more sub-paths in the channel, the worse the performance since more errors will be introduced.

\begin{figure}[t]
	\centering
	\includegraphics[scale=0.4]{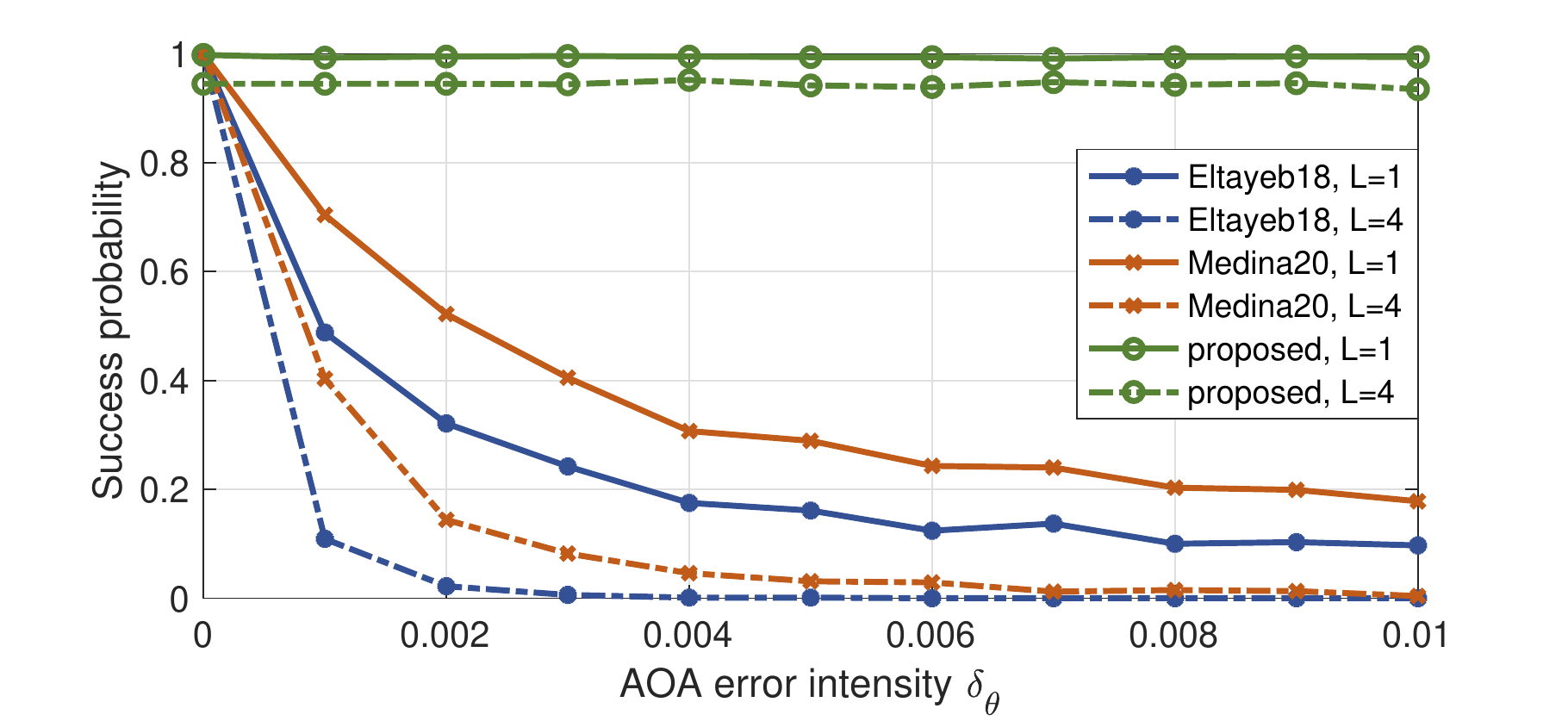}
	\caption{Success probability versus sub-path AOA error intensity under different number of sub-paths.}
	\label{F3}
\end{figure}

Finally, we assume that the estimated AOA $\hat{\theta}_l$ contains estimation error, which is defined as
\begin{equation}
	\hat{\theta}_l = \theta_l + \delta_\theta \theta_\mathrm{e} \pi,
\end{equation}
where $\theta_\mathrm{e} \sim \mathcal{N}(0, 1)$ represents the AOA estimation error and $\delta_\theta$ is the error intensity.
The success probabilities of different techniques under AOA estimation error are shown in Fig. \ref{F3}, where $\mathrm{SNR}=30$dB and the number of measurements is set to $K=N$.
It can be observed that benchmark diagnostic techniques are highly sensitive to AOA estimation errors.
On the contrary, the performance of the proposed diagnostic technique is not affected under all intensities of AOA estimation error since it does not require any CSI knowledge, thereby exhibiting strong robustness against channel estimation errors.

\section{Conclusions}
\label{sec:conclu}
In this letter, we have proposed a blind diagnostic technique for mmWave large-scale antenna systems to locate faulty antenna elements.
By jointly exploiting the sparsity of the mmWave channel and the failure pattern, the location of faulty antenna elements can be identified without any knowledge of the CSI.
A novel atomic norm has been introduced as the sparsity-inducing norm of the mmWave channel, and the diagnosis problem has been formulated as a joint sparse recovery problem.
An efficient ADMM-based diagnostic algorithm has been proposed to solve the joint sparse recovery problem.
Numerical results have shown that the proposed technique has strong robustness against channel estimation errors compared with prior works.

In future works, the diagnosis of mmWave OFDM systems can be considered.
The channels of different subcarriers share a common or similar sparsity pattern, leading to an MMV (multiple measurement vectors) observation model.
By jointly recovering channels and fault deviations of different sub-carriers, the diagnosis performance may be further improved.

\bibliography{ref}

\end{document}